\documentclass[aps,twocolumn,amsmath,letterpaper]{revtex4}
\usepackage{amssymb}
\usepackage{graphicx}
\usepackage{array}
\usepackage{hhline}
\usepackage{longtable}

\begin{document}

    \title{Lower-Critical Spin-Glass Dimension from 23 Sequenced Hierarchical Models}

\author{Mehmet Demirta\c{s}$^{1}$, Asl{\i}  Tuncer$^2$, and A. Nihat Berker$^{1,3}$}
\affiliation{$^1$Faculty of Engineering and Natural Sciences,
Sabanc\i~University, Tuzla 34956, Istanbul, Turkey}
\affiliation{$^2$Department of Physics, Istanbul Technical
University, Maslak 34469, Istanbul, Turkey}
\affiliation{$^3$Department of Physics, Massachusetts Institute of
Technology, Cambridge, Massachusetts 02139, U.S.A.}

\begin{abstract}

The lower-critical dimension for the existence of the Ising
spin-glass phase is calculated, numerically exactly, as $d_L =
2.520$ for a family of hierarchical lattices, from an essentially
exact (correlation coefficent $R^2 = 0.999999$) near-linear fit to
23 different diminishing fractional dimensions. To obtain this
result, the phase transition temperature between the disordered and
spin-glass phases, the corresponding critical exponent $y_T$, and
the runaway exponent $y_R$ of the spin-glass phase are calculated
for consecutive hierarchical lattices as dimension is lowered.


%05.10.Cc    Renormalization group methods
%64.60.ae    Renormalization-group theory
%64.60.De    Statistical mechanics of model systems
%75.10.Nr    Spin-glass and other random models
%75.50.Lk    Spin glasses and other random magnets

\end{abstract}

    \maketitle
    \def\s{\rule{0in}{0.28in}}
    \setlength{\LTcapwidth}{\columnwidth}

\section{Introduction}

Singular phase diagram behavior as a function of spatial
dimensionality $d$ compounds the interest and challenge of the phase
transitions problems, as effectively pausing the "phase transition
of phase transitions" problem. Most visible are the lower-critical
dimensions, which are the spatial dimensional thresholds for
different types of orderings. For example, the lower-critical
threshold for ferromagnetic ordering in magnetic systems is $d_L=1$
for one-component (Ising) spins and $d_L=2$ for spins with more than
one component.  Similarly, the lower-critical dimensions for
ferromagnetic ordering under quenched random fields
\cite{Jaccarino,Birgeneau,Wong,Berker84,Aizenman,AizenmanE,Falicov}
are respectively $d_L=2$ and $d_L=4$, for one-component spins and
for spins with more than one component.

A very recent experimental study \cite{Orbach} on Ge:Mn films has
shown the spin-glass lower-critical temperature to be $2<d_L<3$.
This is consistent with the earlier theoretical result of $d_L=2.5$
from replica symmetry-breaking mean-field theory.\cite{Parisi} A
numerical fit to the spin-glass critical temperatures for integer
dimensions has also suggestted $d_L=2.5$.\cite{Boettcher} Other
theoretical work have claimed $d_L=4$ from earlier ordered-phase
stability studies \cite{Moore1,Sompolinsky,Dedominicis}, $2<d_L<3$
from transfer-matrix studies \cite{Moore2},and more recently $d_L=2$
from Monte Carlo \cite{Matsubara,Houdayer} and ground-state studies
\cite{Young}. Renormalization-group work, on in effect two
hierarchical lattices \cite{Southern} different from ours have
earlier obtained $2<d_L<3$ and on a family of hierarchical lattices
\cite{Amoruso}, again different from ours, find $d_L$ close to 2.5.

The lower-critical dimensions need not be integer, in view of
physical systems on fractal/hierarchical lattices and algebraic
manipulations that analytically continue.  In fact, it would be
highly interesting to find a lower-critical dimension that is
neither an integer, nor a simple fraction. Our current study
indicates that this is in fact the case for the family of
hierarchical lattices studied here, with $d_L=2.520$. We obtain this
result from a remarkably good fit to the renormalization-group
runaway exponent $y_R$ from the numerically exact
renormalization-group solution of a family of 23 hierarchical models
with non-integer dimensions $d =$ 2.46, 2.63, 2.77, 2.89, 3.00,
3.10, 3.18, 3.26, 3.33, 3.40, 3.46, 3.52, 3.58, 3.63, 3.68, 3.72,
3.77, 3.81, 3.85, 3.89, 3.93, 3.97, 4.00. Our result is also
consistent with the results that are graphically displayed in
Ref.\cite{Amoruso} for a \underline{different} family of
hierarchical lattices.

\section{Lower-Critical Dimension from Sequenced Hierarchical Models}

Hierarchical models are constructed
\cite{BerkerOstlund,Kaufman1,Kaufman2,McKay,Hinczewski1} by
imbedding a graph into a bond, as examplified in Fig. 1, and
repeating this procedure by self-imbedding infinitely many times.
This procedure can also be done on units with more than two external
vertices, \textit{e.g.}, the layered Sierpinski gasket in Ref.
\cite{BerkerMcKay}. When interacting systems are placed on
hierarchical lattices, their renormalization-group solution proceeds
in the reverse direction than the lattice build-up just described,
each eliminated elementary graph generating a renormalized
interaction strength for the ensuing elementary bond. Hierarchical
lattices were originally introduced \cite{BerkerOstlund} as
presenting exactly soluble models with renormalization-group
recursion relations that are identical to those found in approximate
position-space renormalization-group treatments of Euclidian
lattices \cite{Migdal,Kadanoff}, identifying the latter as
physically realizable approximations. However, from the above, it is
clear that any graph (or graphs \cite{Hinczewski1}) may be chosen in
the self-imbedding procedure and one need not be faithful to any
approximate renormalization-group solution. Hierarchical lattices
have been used to study a variety of spin-glass and other
statistical mechanics
problems.\cite{Gingras2,Migliorini,Gingras1,Hinczewski,Guven,Ohzeki,Ozcelik,Gulpinar,Ilker1,Ilker2,Ilker3,Kaufman,
Barre, Monthus,
Zhang,Shrock,Xu,Herrmann1,Herrmann2,Hwang2013,Garel,Hartmann,Fortin,Wu,Timonin,Derrida,Thorpe,Hasegawa,Monthus2,Lyra,Singh,
Xu2014,Hirose}

The length rescaling factor $b$ in a hierarchical lattice is the
number of bonds on the shortest distance between the external
vertices of the elementary graph which is replaced by a single bond
in one scale change. The volume rescaling factor $b^d$ is the number
of bonds inside the elementary graph.  From these two rescaling
factors, the dimensionality $d$ is extracted, as exemplified in Fig.
1.  In our study, $b=3$ is used in order to treate the ferromagnetic
and antiferromagnetic correlations on equal footing. The
lower-critical dimension of spin-glass systems is studied here by
considering a systematic family of hierarchical lattices in all its
possible decreasing dimensions.

\section{The spin-glass system and the renormalization-group method}

The Ising spin-glass system is defined by the Hamiltonian
\begin{equation}
-\beta \mathcal{H}=\sum_{\langle ij \rangle} J_{ij} s_i s_j
\end{equation}
where $\beta=1/kT$, at each site $i$ of a lattice the spin $s_i =
\pm 1$, and $\langle ij \rangle$ denotes that the sum runs over all
nearest-neighbor pairs of sites. The bond strengths $J_{ij}$ are
$+J>0$ (ferromagnetic) with probability $1-p$ and $-J$
(antiferromagnetic) with probability $p$.

\begin{figure}
\includegraphics[scale=0.85]{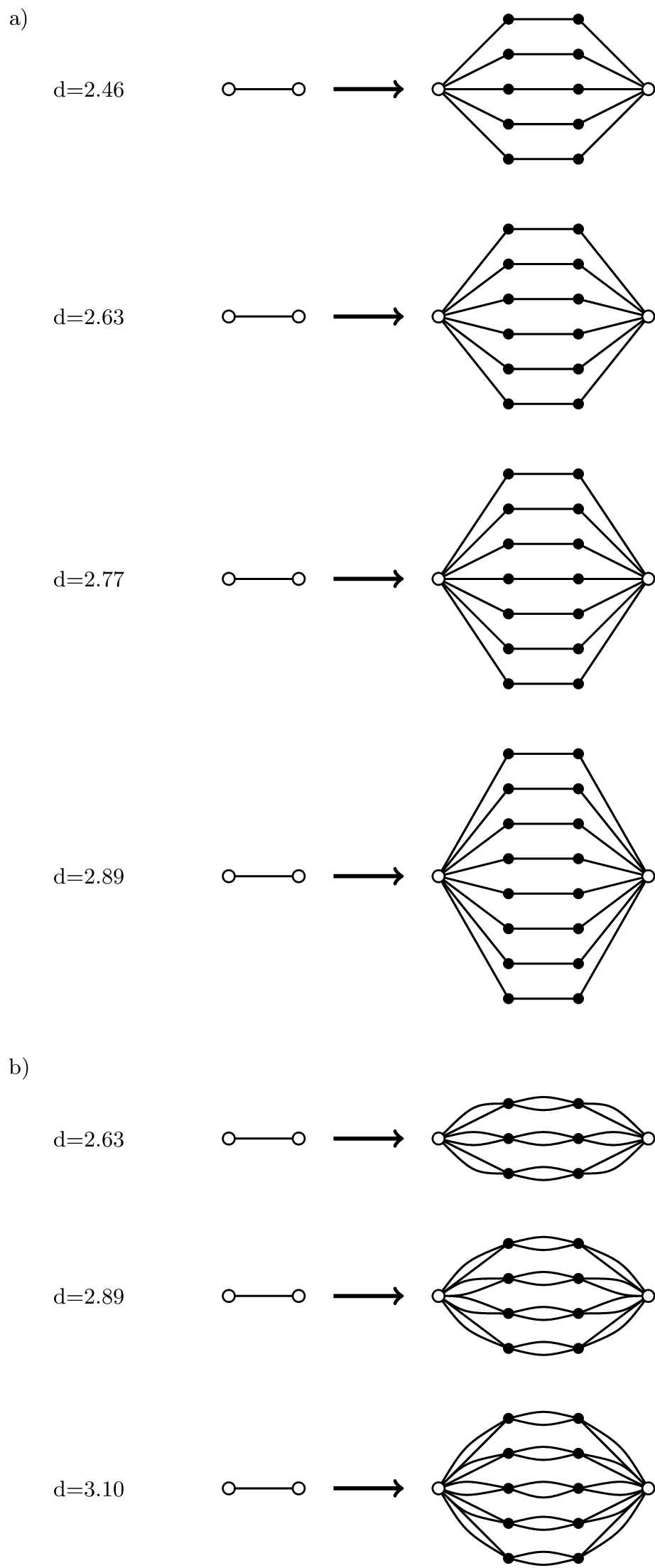}
\caption{(a) The construction of the family of hierarchical lattices
used in this study.  Each lattice is constructed by repeatedly
self-imbedding the graph.  The graphs here are $n$ parallel series
of $b=3$ bonds.  The dimension $d = 1 + \ln n / \ln b$ of each
lattice is given. The renormalization-group solution consists in
implementing this process in the reverse direction, for the
derivation of the recursion relations of the local interactions. The
lattices shown here and 19 other lattices with the nearby fractional
dimensions are used in our calculations. (b) The family of
hierarchical lattices with $n_1$ parallel $b=3$ series of $n_2$
parallel bonds. The resulting hierarchical models are equivalent to
the family in (a) with $n = n_1 n_2$, with respect to identical
critical exponents and phase diagram topology including the
occurrence/nonoccurrence of a spin-glass phase.}
\end{figure}

The renormalization-group transformation is achieved by a
decimation,
\begin{equation}
e^{J_{im}^{(dec)}s_i s_m + G_{im}}=\sum_{s_j,s_k} e^{J_{ij} s_i
s_j+J_{jk} s_j s_k +J_{km} s_k s_m},
\end{equation}
where the additive constants $G_{ij}$ are unavoidably generated,
followed by $n$ bond movings,
\begin{equation}
J_{ij}^{(bm)} = \sum_{k=1}^n J_{ij}^{(k)}.
\end{equation}
\begin{figure}[]
\centering
\includegraphics[scale=1]{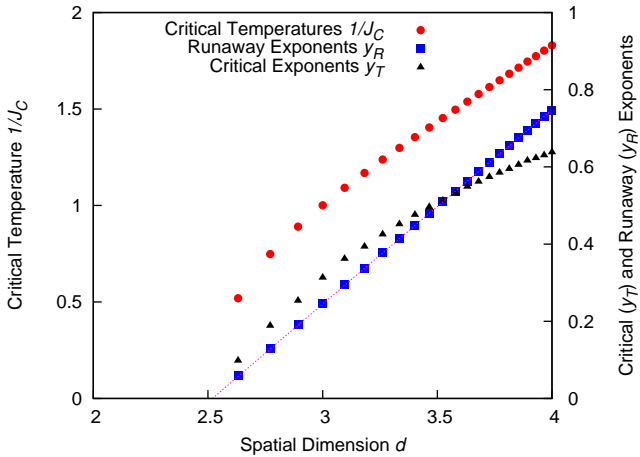}
\caption{(Color online) Critical temperatures $1/J_c$ and critical
exponents $y_T$ of the phase transitions between the spin-glass and
paramagnetic phases as a function of dimension $d$, for the
hierarchical models with antiferromagnetic bond concentration
$p=0.5$.  The runaway exponents $y_R$ of the spin-glass phase are
also shown and give a perfect fit to $y_R = -1.30908 + 0.528513 d -
0.00354805 d^2$, leading with a small extrapolation to the
lower-critical dimension $d = 2.520$ for $y_R = 0$, with a very
satisfactory correlation coefficient of $R^2 = 0.999999$.}
\end{figure}
\begin{table}[]
\centering \rule{0pt}{2.6ex}  \renewcommand{\arraystretch}{1.3}
\begin{ruledtabular}
    \begin{tabular}{cccc}
    Spatial & Critical & Critical & Runaway \\
    Dimension & Temperatures & Exponents & Exponents \\
    $d$ & $1/J_C$ & $y_T$ & $y_R$ \\
    \hline
    2.630930 & 0.519268 & 0.098077 & 0.058731 \\
    2.771244 & 0.747982 & 0.188596 & 0.129983 \\
    2.892789 & 0.890503 & 0.253690 & 0.191904 \\
    3.000000 & 1.001319 & 0.313414 & 0.246144 \\
    3.095903 & 1.091770 & 0.361975 & 0.294649 \\
    3.182658 & 1.168653 & 0.393837 & 0.338155 \\
    3.261860 & 1.237723 & 0.425397 & 0.377881 \\
    3.334718 & 1.298225 & 0.451743 & 0.414440 \\
    3.402174 & 1.354258 & 0.476199 & 0.448214 \\
    3.464974 & 1.404661 & 0.495999 & 0.479850 \\
    3.523719 & 1.452817 & 0.513016 & 0.509181 \\
    3.578902 & 1.496452 & 0.531699 & 0.536880 \\
    3.630930 & 1.538271 & 0.549022 & 0.563149 \\
    3.680144 & 1.577300 & 0.562079 & 0.587707 \\
    3.726833 & 1.613844 & 0.573932 & 0.610941 \\
    3.771244 & 1.649036 & 0.585283 & 0.633434 \\
    3.813588 & 1.682659 & 0.594959 & 0.654932 \\
    3.854050 & 1.714417 & 0.605789 & 0.675080 \\
    3.892789 & 1.745469 & 0.616496 & 0.693914 \\
    3.929947 & 1.774176 & 0.623179 & 0.712461 \\
    3.965647 & 1.802906 & 0.630527 & 0.730927 \\
    4.000000 & 1.829792 & 0.638313 & 0.747294 \\
    \end{tabular}%
    \end{ruledtabular}
\rule[-1.2ex]{0pt}{0pt}%
\caption{Critical temperatures $1/J_c$ and critical exponents $y_T$
of the phase transitions between the spin-glass and paramagnetic
phases as a function of dimension $d$, for the hierarchical models
with antiferromagnetic bond concentration $p=0.5$.  The runaway
exponents $y_R$ of the spin-glass phase are also shown and give a
perfect fit to $y_R = -1.30908 + 0.528513 d - 0.00354805 d^2$,
leading with a small extrapolation to the lower-critical dimension
$d = 2.520$ for $y_R = 0$, with a very satisfactory correlation
coefficient of $R^2 = 0.999999$.} \label{tab:addlabel}
\end{table}
The starting bimodal quenched probability distribution of the
interactions, characterized by $p$ and described above, is not
conserved under rescaling. The renormalized quenched probability
distribution of the interactions is obtained by the convolution
\cite{Andelman}
\begin{equation}
P'(J'_{i'j'}) = \int{\left[\prod_{ij}^{i'j'}dJ_{ij}
P(J_{ij})\right]} \delta(J'_{i'j'}-R(\left\{J_{ij}\right\})),
\end{equation}
where $R(\left\{J_{ij}\right\})$ represents the decimation and bond
moving given in Eqs.(2) and (3).  For numerical practicality, the
bond moving and decimation of Eqs.(2) and (3) are achieved by a
sequence of pairwise combinations of interactions, each pairwise
combination leading to an intermediate probability distribution
resulting from a pairwise convolution as in Eq.(4). The probability
distribution is represented by 200 histograms
\cite{Migliorini,Hinczewski,Guven,Ozcelik,Gulpinar,Ilker2}, which
are apportioned in $J\gtrless 0$ according to total probability
weight. The histograms are distributed in the interval $J_+ \pm 2.5
\sigma_+$, where  $J_+$ and $\sigma_+$ are the average and standard
deviation of the $J>0$ interactions, and similarly for the $J<0$
interactions.

The different thermodynamic phases of the system are identified by
the different asymptotic renormalization-group flows of the quenched
probability distributions.  For all renormalization-group flows,
inside the phases and on the phase boundaries, Eq.(4) is iterated
until asymptotic behavior is reached. Thus, we are able to calculate
phase transition temperatures and, by linearization around the
unstable asymptotic fixed distribution of the phase boundaries,
critical exponents. Similar previous studies, on other spin-glass
systems, are in Refs.
\cite{Gingras2,Migliorini,Gingras1,Hinczewski,Guven,Ohzeki,Ozcelik,Gulpinar,Ilker1,Ilker2}.

\section{Diminishing Critical, Runaway Exponents, Critical Temperatures and the Lower-Critical Dimension of the Sequence}

For our chosen sequence of hierarchical systems (Fig. 1), we have
calculated, at antiferromagnetic bond concentration $p=0.5$, the
phase transition temperature $1/J_C$ where the renormalization-group
flows bifurcate between the disordered-phase and the
spin-glass-phase attractor sinks.  The spin-glass sink is
characterized by an interaction probability distribution $P(J_{ij})$
that is symmetric in ferromagnetism-antiferromagnetism
$(J_{ij}\gtrless 0)$ and that diverges in interaction absolute
value: The average interaction strength $<|J|>$ across the system
diverges as $b^{n y_R}$ where $n$ is the number of
renormalization-group iterations and $y_R > 0$ is the runaway
exponent.  The spin-glass sink and simultaneously the spin-glass
phase disappears when the runaway exponent $y_R$ reaches
0.\cite{Ilker2}  The calculated spin-glass phase transition
temperatures, critical and runaway exponents are given in Fig. 2 and
in Table I as a function of spatial dimension $d$. The lattice with
$d=2.46$, not having a spin-glass phase, is below the lower-critical
dimension. For the 22 other consecutive lattices with a spin-glass
phase, we have chosen to fit the runaway exponent values, since they
gives an excellent, near-linear fit with
\begin{equation}
y_R = -1.30908 + 0.528513 d - 0.00354805 d^2,
\end{equation}
with an amazingly satisfactory correlation coefficient of $R^2 =
0.999999$. This fit gives, with a small extrapolation, $y_R = 0$ for
$d = 2.520$. Note the near linearity, namely the smallness of the
quadratic coefficient in Eq. (5). (In fact, a linear fit gives $y_R
= 0$ for $d = 2.516$, with a little less amazingly satisfactory
correlation coefficient of $R^2 = 0.999992$.)

Our calculated lower-critical dimension $d_L$, where the spin-glass
phase disappears at zero-temperature, is thus seen to be $d_L =
2.520$, for the sequence of hierarchical lattices studied here. It
is noteworthy that $d_L$ is not an integer and not even a simple
fraction, contrary previous examples of lower-critical dimensions
(and even contrary to upper-critical dimensions, where mean-field
behavior sets in) for other models.

Another important quantity is the critical exponent $y_T = 1/\nu >
0$ of the phase transition between the disordered and spin-glass
phases. This exponent is calculated from the scaling behavior of
small deviations of the average interaction strength from its fixed
finite value at the unstable fixed distribution of the phase
transition. The calculated critical exponents are also given in Fig.
2.  As the spatial dimension is lowered, $y_T$ also approaches 0.
At the lower-critical dimension, $y_T$ reaches 0.  The
disordered-spin-glass phase transition disappears at $d_L$, where
the spin-glass phase disappears.

\section{Conclusion}

Our family of hierarchical lattices (Fig. 1) yields smooth and
systematic behavior in all three quantities: the critical
temperatures $1/J_C$, the critical exponents $y_C$, and, eminently
fitably, the runaway exponents $y_R$.  All three quantities yield
the lower-critical temperature of $d_L = 2.520$. It is noteworthy
that $d_L$ is not an integer and not even a simple fraction,
contrary previous examples of lower-critical dimensions (and even
contrary to upper-critical dimensions, where mean-field behavior
sets in) for other models.

\begin{acknowledgments}
We thank Prof. H. Nishimori for suggesting this calculation to us.
Support by the Alexander von Humboldt Foundation, the Scientific and
Technological Research Council of Turkey (T\"UBITAK), and the
Academy of Sciences of Turkey (T\"UBA) is gratefully acknowledged.
\end{acknowledgments}

\end{document}